\documentclass[prc,aps,floatfix,showpacs,twocolumn]{revtex4}
\usepackage{dcolumn}
\usepackage{bm}
\usepackage{graphicx}
\begin{document}

\title{Dependence of two-nucleon momentum densities on total pair momentum}
\author{R.B.\ Wiringa$^1$, R.\ Schiavilla$^{2,3}$, Steven C.\ Pieper$^1$, and J.\ Carlson$^4$}
\affiliation{
$^1$Physics Division, Argonne National Laboratory, Argonne, IL 60439\\
$^2$Theory Center, Jefferson Laboratory, Newport News, VA 23606 \\
$^3$\mbox{Department of Physics, Old Dominion University, Norfolk, VA 23529}\\
$^4$\mbox{Theoretical Division, Los Alamos National Laboratory, Los Alamos, NM 87545}
}

\date{\today}

\begin{abstract}
Two-nucleon momentum distributions are calculated for the ground states of
$^3$He and $^4$He as a function of the nucleons' relative and total momenta.
We use variational Monte Carlo wave functions derived from a realistic 
Hamiltonian with two- and three-nucleon potentials.  
The momentum distribution of $pp$ pairs is found to be much smaller than 
that of $pn$ pairs for values of the relative momentum in the
range (300---500) MeV/c and vanishing total momentum.  
However, as the total momentum increases to 400 MeV/c, the ratio of
$pp$ to $pn$ pairs in this relative momentum range grows and approaches
the limit 1/2 for $^3$He and 1/4 for $^4$He, corresponding to the ratio
of $pp$ to $pn$ pairs in these nuclei.  This behavior should be easily
observable in two-nucleon knock-out processes, such as $A(e,e^\prime pN)$.
\end{abstract}


\maketitle
In a recent letter, we studied the role of tensor forces on the 
correlations between pairs of nucleons in light nuclei~\cite{Schiavilla07}. 
In that work we reported calculations of the relative momentum distribution
of $pp$ and $pn$ pairs with vanishing total momentum.
We found that the strong spatial-spin-isospin correlations induced by
the tensor force lead to large differences in the $pp$ and $pn$ distributions
at moderate values of the relative momentum in the pair.
These differences have been observed in a two-nucleon knockout experiment
on $^{12}$C at Jefferson Laboratory (JLab)~\cite{Subedi08}.

In this note, we report an extension of our calculations to finite
total momentum of the correlated pair for $^3$He and $^4$He nuclei.
This is motivated by a preliminary analysis of data on $^3$He from the CEBAF
large acceptance spectrometer (CLAS) collaboration at JLab~\cite{Weinstein08}.
We find that the large differences in $pp$ and $pn$ distributions gradually
diminish as the center-of-mass momentum increases, until it approaches
the ratio of $pp$ to $pn$ pairs for a given whole nucleus.

The probability of finding two nucleons with relative momentum ${\bf q}$
and total momentum ${\bf Q}$ in isospin state $TM_T$ in the
ground state of a nucleus is proportional to the density
\vspace*{-.2in}
\begin{widetext}
\vspace*{-.2in}
\begin{eqnarray}
\rho_{TM_T}({\bf q},{\bf Q})\!\!&=&\!\!\frac{A(A-1)}{2\, (2J+1)}\! \sum_{M_J}
\int d{\bf r}_1\, d{\bf r}_2\, d{\bf r}_3 \cdots d{\bf r}_A\, 
d{\bf r}^\prime_1\,d{\bf r}^\prime_2 \,
\psi^\dagger_{JM_J}({\bf r}_1^\prime,{\bf r}_2^\prime,{\bf r}_3, \dots,{\bf r}_A)\,  \nonumber \\
&& \times \, e^{-i{\bf q}\cdot ({\bf r}_{12}-{\bf r}_{12}^\prime)} \,
   e^{-i{\bf Q}\cdot ({\bf R}_{12}-{\bf R}_{12}^\prime)}
\, P_{TM_T} (12) \, 
\psi_{JM_J} ({\bf r}_1,{\bf r}_2,{\bf r}_3, \dots,{\bf r}_A) \ ,
\label{eq:rhoqQ}
\end{eqnarray}
\end{widetext}
\vspace*{-.2in}
where 
${\bf r}_{12}\equiv {\bf r}_1-{\bf r}_2$,
${\bf R}_{12}\equiv ({\bf r}_1+{\bf r}_2)/2$, and similarly for
${\bf r}^\prime_{12}$ and ${\bf R}^\prime_{12}$.  Here
$P_{TM_T}(12)$
is the isospin projection operator, and $\psi_{JM_J}$ denotes
the nuclear wave function in spin and spin-projection state
$JM_J$.  The normalization is
\begin{equation}
\int \frac{d{\bf q}}{(2\pi)^3}
 \frac{d{\bf Q}}{(2\pi)^3}\, \rho_{TM_T}({\bf q},{\bf Q})=
N_{TM_T} \ ,
\end{equation}
where $N_{TM_T}$ is the number of $N\!N$ pairs in state
$TM_T$.  
Obviously, integrating $\rho_{TM_T}({\bf q},{\bf Q})$
over only ${\bf Q}$ gives the probability of finding two nucleons with
relative momentum ${\bf q}$, regardless of their pair momentum ${\bf Q}$
(and vice-versa).

For this study we use variational Monte Carlo (VMC) wave functions,
derived from a realistic Hamiltonian consisting of the Argonne $v_{18}$
two-nucleon~\cite{Wiringa95} and Urbana-IX three-nucleon~\cite{Pudliner95}
interactions (AV18/UIX).
The double Fourier transform in Eq.~(\ref{eq:rhoqQ}) is computed by Monte
Carlo (MC) integration.  A standard Metropolis walk, guided by 
$|\psi_{JM_J} ({\bf r}_1,{\bf r}_2,{\bf r}_3, \dots,{\bf r}_A)|^2$, is used
to sample configurations~\cite{Pieper01}.  For each configuration 
a two-dimensional grid of
Gauss-Legendre points, $x_i$ and $X_j$, is used to compute the Fourier transform.
Instead of just moving the $\psi^\prime$ position (${\bf r}_{12}^\prime$ and  
${\bf R}_{12}^\prime$) away from a fixed $\psi$ position (${\bf r}_{12}$ and  
${\bf R}_{12}$), both positions are moved symmetrically away from
${\bf r}_{12}$ and  ${\bf R}_{12}$, so Eq.~(\ref{eq:rhoqQ}) becomes
\vspace*{-.2in}
\begin{widetext}
\vspace*{-.2in}
\begin{eqnarray}
\rho_{TM_T}({\bf q},{\bf Q}) = \frac{A(A-1)}{2\, (2J+1)}\! \sum_{M_J}
\int d{\bf r}_1\, d{\bf r}_2\, d{\bf r}_3 \cdots d{\bf r}_A \, d{\bf x}\, d{\bf X}\, 
\psi^\dagger_{JM_J} 
({\bf r}_{12}\!+\!{\bf x}/2,{\bf R}_{12}\!+\!{\bf X}/2,{\bf r}_3, \dots,{\bf r}_A)&&  \nonumber \\
\times  \,  e^{-i{\bf q}\cdot {\bf x} } \, e^{-i{\bf Q}\cdot {\bf X} } \, P_{TM_T} (12) \,
\psi_{JM_J} 
({\bf r}_{12}\!-\!{\bf x}/2,{\bf R}_{12}\!-\!{\bf X}/2,{\bf r}_3, \dots,{\bf r}_A)&& .
\end{eqnarray}
\end{widetext}
\vspace*{-.2in}
Here the polar angles of the $x$ and $X$ grids are also sampled by MC 
integration, with one sample per pair.
This procedure is similar to that adopted most recently in studies of the
$^3$He$(e,e^\prime p)d$ and $^4$He$(\vec{e},e^\prime \vec{p}\,)^3$H
reactions~\cite{Schiavilla05}, and has the advantage of very substantially
reducing the statistical errors originating from the rapidly oscillating
nature of the integrand for large values of $q$ and $Q$.  

The present method is computationally intensive, 
because complete Gaussian integrations have to be performed for each of
the configurations sampled in the random walk.
The large range of values of $x$ and $X$ required to obtain converged 
results,  especially for $^3$He, require fairly large numbers of points; 
we used grids of up to 96 and 80 points for $x$ and $X$, respectively.
We also sum over all pairs in a given MC sample instead of just a single pair.

\begin{figure}[b!]
\includegraphics[angle=-90,width=3.25in]{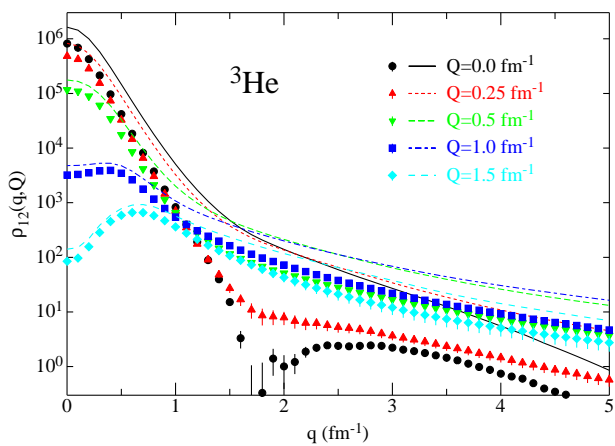}
\caption{(Color online) The $pn$ (lines) and $pp$ (symbols) momentum 
distributions in $^3$He as functions of the relative momentum 
$q$ at total pair momentum $Q$ from 0 to 2 fm$^{-1}$.}
\label{fig:plot3}
\end{figure}
 
\begin{figure}[b!]
\includegraphics[angle=-90,width=3.25in]{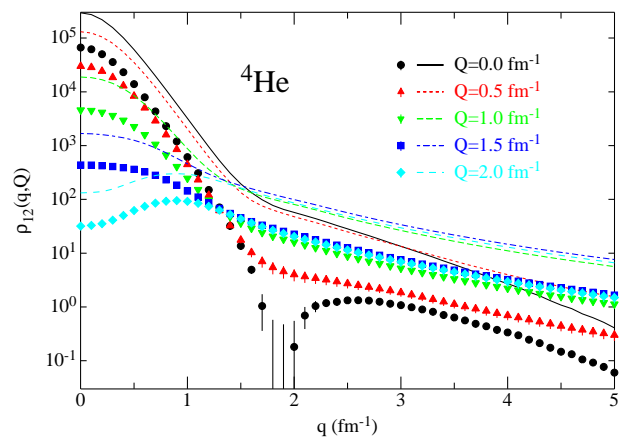}
\caption{(Color online) The $pn$ (lines) and $pp$ (symbols) momentum 
distributions in $^4$He as functions of the relative momentum 
$q$ at total pair momentum $Q$ from 0 to 2 fm$^{-1}$.}
\label{fig:plot4}
\end{figure}

The $pn$ and $pp$ distributions at five values of the total momentum
$Q$, with ${\bf Q} \parallel {\bf q}$, are shown as functions of 
the relative momentum $q$ for $^3$He in Fig.~\ref{fig:plot3} and
for $^4$He in Fig.~\ref{fig:plot4}.
The statistical errors due to the MC integration are displayed 
only for the $pp$ pairs; they are comparable for the $pn$ pairs.
When the total momentum vanishes, there is a node in the $pp$ relative
momentum distribution just below 2 fm$^{-1}$, while the $pn$ distribution
has a broad shoulder in this region.
Integration over the relative momenta in the range 1.5---2.5 fm$^{-1}$
gives a ratio of $pp$ to $pn$ pairs $R_{pp/pn} = 0.014 \pm 0.004$ for $^3$He,
compared to 1/2 for the whole nucleus integrated over all ${\bf q}$ 
and ${\bf Q}$.
For $^4$He and $Q$=0 the value is $R_{pp/pn} = 0.023 \pm 0.006$,
compared to 1/4 for the whole nucleus.

The much greater magnitude of the $pn$ momentum distribution is due to the
strong correlations induced by tensor components in the underlying
$N\!N$ potential.  
When $Q$=0, the pair and residual $(A-2)$ system are in a relative $S$-wave.
Hence, in $^3$He ($^4$He), whose spin-parity is $\frac{1}{2}^+$ (0$^+$),
$pn$ pairs are predominantly in $T$=0 and $^3$S$_1$-$^3$D$_1$ (deuteron-like)
states, while $pp$ pairs are in $T$=1 and $^1$S$_0$ (quasi-bound)
states~\cite{Forest96}.  
The $D$-wave component of the deuteron-like pairs fills in the 
node in the $S$-wave momentum distribution.

\begin{figure}[bth]
\includegraphics[angle=-90,width=3.25in]{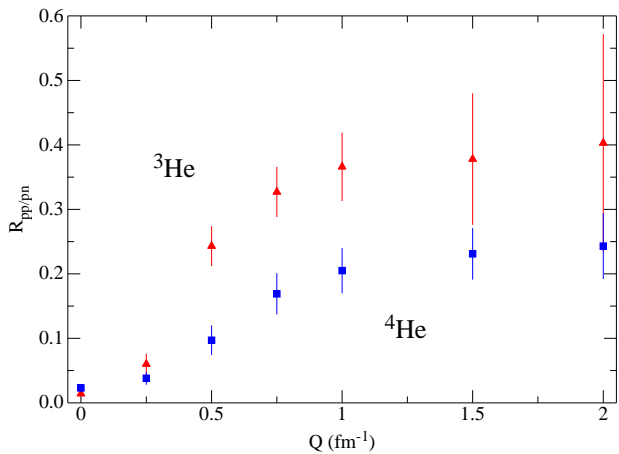}
\caption{(Color online) The ratio of $pp$ to $pn$ pairs integrated over
relative momentum ${\bf q} \parallel {\bf Q}$ of 1.5 to 2.5 fm$^{-1}$ 
as a function of total 
momentum $Q$; red triangles are for $^3$He and blue squares for $^4$He.}
\label{fig:ratio}
\end{figure}

However, for $Q>0$, the two clusters may have non-zero orbital 
angular momentum and hence the $pp$ and $pn$ pairs are no longer
constrained to be in the quasi-bound or deuteron-like states.
Thus the $S$-wave node in the $pp$ pairs can be filled in by higher
angular momentum states.
Figures ~\ref{fig:plot3} and \ref{fig:plot4} show that this does happen;
the node in the $pp$ relative momentum distribution is rather rapidly
filled in as $Q$ increases.
Consequently $R_{pp/pn}$ increases as $Q$ increases, as
shown in Fig.~\ref{fig:ratio} for $q$ integrated over 1.5---2.5 fm$^{-1}$.
 
The most direct evidence for tensor correlations in nuclei comes from 
measurements of the deuteron structure functions and tensor polarization 
by elastic electron scattering~\cite{Alexa99}.  
In essence, these measurements have mapped out the Fourier transforms 
of the charge densities of the deuteron in states with spin
projections $\pm1$ and $0$, showing that they are very different.  
In other processes, such as $^2$H$(d,\gamma)^4$He~\cite{Arriaga91} 
at very low energy, or proton knock-out from a polarized deuteron~\cite{Zhou99}
the effects of tensor correlations are more subtle and their presence is not 
easily isolated in the experimental data.  This is because of corrections
from initial or final state interactions and many-body terms in the
transition operators.

Some of these corrections will also affect, for instance, the
cross sections for $(e,e^\prime pn)$ and $(e,e^\prime pp)$
knock-out processes in back-to-back kinematics.  However,
one would expect the contributions due to final state
interactions in the $pn$ and $pp$ reactions, both between the
nucleons in the pair and between these and the nucleons in
the residual $(A-2)$ system, to be of similar magnitude for
relative momenta in the range (300---500) MeV/c.  In particular,
charge-exchange processes have been estimated to give small ($\le 10$\%)
corrections to these ratios in $^{12}$C at JLab kinematics~\cite{Subedi08}.
Such processes, which are induced by interactions between
the knocked-out pair and the residual cluster, could change
an initial $p$$n$ pair on its way out of the nucleus into
the (detected) $p$$p$ pair, thus increasing the $(e,e^\prime pp)$
to $(e,e^\prime pn)$ cross section ratios.  Lastly, leading
terms in the electromagnetic two-body current vanish in $pp$
because of their isospin structure~\cite{Marcucci05}.  Of
course, they will contribute in $pn$, but are not expected
to produce large effects.

The recent experiment at JLab referred to earlier has measured the ratio of
$^{12}$C$(e,e^\prime pn)$ to $^{12}$C$(e,e^\prime pp)$
cross sections in back-to-back kinematics for relative
momenta in the range 300---500 MeV/c~\cite{Subedi08} to be 
$\simeq 10$.  These measurements have corroborated the
results of an earlier analysis of a BNL experiment, which
measured cross sections for $(p,pp)$ and $(p,ppn)$ processes
on $^{12}$C in similar kinematics~\cite{Piasetzky06a}.
The observed enhancement in the $p$$n$ to $p$$p$ ratio
is in agreement with the prediction of Ref.~\cite{Schiavilla07},
and beautifully demonstrates the crucial role that
the tensor force plays in shaping the short-range
structure of nuclei.

It would be interesting to extend these measurements to other nuclei.
In $^3$He and $^4$He, one would expect the node in the $p$$p$ momentum
distribution to be filled in by interaction effects in the final
state~\cite{Schiavilla05}.  However, the ratio of $p$$p$ to $p$$n$ cross
sections in the range (300---500) MeV/c should still reflect the dominance
of the $p$$n$ momentum distribution at these values of relative
momenta.  In fact, the analysis of JLab CLAS data on $^3$He mentioned above
suggests that this is indeed the case~\cite{Weinstein08}.  These data
also seem to confirm the rapid rise of the $p$$p$ to $p$$n$ ratio with
increasing total pair momentum, predicted in Fig.~\ref{fig:ratio}
of the present work.

\section*{Acknowledgments}

A stimulating conversation with L. Weinstein
is gratefully acknowledged by one of the authors (R.S.).
This work is supported by the U.S.\ Department of Energy, Office of
Nuclear Physics, under contracts DE-AC02-06CH11357 (R.B.W.\ and S.C.P.),
DE-AC05-06OR23177 (R.S.), and DE-AC52-06NA25396 (J.C.).  
The calculations were made at Argonne's Laboratory Computing Resource Center.
\end{document}